\documentstyle[aps,multicol]{revtex}

\begin{document}
\draft

\newcommand{\beq}{\begin{equation}} 
\newcommand{\eeq}{\end{equation}}
\newcommand{\bqa}{\begin{eqnarray}} 
\newcommand{\eqa}{\end{eqnarray}}
\newcommand{\nn}{\nonumber} 
\newcommand{\erf}[1]{Eq.~(\ref{#1})}
\newcommand{\dg}{^\dagger}
\newcommand{\smallfrac}[2]{\mbox{$\frac{#1}{#2}$}}
\newcommand{\bra}[1]{\langle{#1}|} 
\newcommand{\ket}[1]{|{#1}\rangle}
\newcommand{\ip}[1]{\left\langle{#1}\right\rangle}
\newcommand{\sch}{Schr\"odinger } 
\newcommand{\schs}{Schr\"odinger's }
\newcommand{\hei}{Heisenberg } 
\newcommand{\heis}{Heisenberg's }
\newcommand{\half}{\smallfrac{1}{2}} 
\newcommand{\bl}{{\bigl(}}
\newcommand{\br}{{\bigr)}} 
\newcommand{\ve}{\varepsilon}
\newcommand{\rt}[1]{\sqrt{#1}\,}

\title{In-loop squeezing is real squeezing to an in-loop atom}
\author{H.\ M.\ Wiseman\protect\cite{*}}
\address{Department of Physics, The University of Queensland, 
St.\ Lucia 4072, Australia}

\date{\today}
 
\maketitle

\begin{abstract}
	
Electro-optical feedback can produce an in-loop 
photocurrent with arbitrarily low noise.  This is not regarded as 
evidence of `real' squeezing because squeezed light cannot be 
extracted from the loop using a linear beam splitter.  Here I show 
that illuminating an atom (which is a nonlinear optical 
element) with `in-loop' squeezed light causes line-narrowing of one quadrature of 
the atom's fluorescence.  This has long been regarded as an effect 
which can only be produced by squeezing.  Experiments on atoms using 
in-loop squeezing should be much easier than those with conventional 
sources of squeezed light.

\end{abstract}   

\pacs{42.50.Dv, 42.50.Ct, 42.50.Lc, 32.80.-t}

\begin{multicols}{2}

Squeezing is the reduction in noise in one quadrature of a light beam 
below the standard quantum limit at the expense of a corresponding 
increase in the noise of the conjugate quadrature \cite{WalMil94}. 
The characteristic below and above shot-noise homodyne photocurrent 
spectra were first observed by Slusher {\em et al.} in 1985\cite{Slu85}. 
Around the same time, the production of a below shot-noise 
amplitude-quadrature spectrum for a photocurrent which was part of a 
control loop was also first reported \cite{WalJak85,MacYam86}.  As 
pointed out by Shapiro {\em et al.}\cite{Sha87}, this cannot be taken 
as evidence for squeezing in the conventional sense because the two-time
commutation relations for an in-loop field are not those of a free 
field.  Moreover, attempts to remove some of the supposedly 
low-noise light by a beam splitter yielded only above shot-noise light 
\cite{WalJak85,YamImoMac86}.

Very soon after the first observation of squeezing, Gardiner 
\cite{Gar86} made a seminal prediction regarding 
its effects on matter \cite{FicDru97}, namely that 
 immersing an atom in broad-band squeezed light would break the 
equality between the transverse decay rates for the two quadratures of 
the atomic dipole.  
In particular, one decay rate could be made arbitrarily small, 
producing an arbitrarily narrow line in the power spectrum of the 
atom's fluorescence.  This was seen, as the title of Ref.~\cite{Gar86} 
proclaims, as a ``direct effect of squeezing''.  

In this letter I pose the question of whether this atomic 
line-narrowing is characteristic only of squeezing in the conventional 
sense (`real squeezing'), or whether it can be produced by light which 
gives rise to a below shot-noise photocurrent by virtue of being 
part of a feedback loop (`in-loop squeezing').  The answer is that 
in-loop squeezing {\em can} do the job. In fact, 
the dependence of the line-narrowing on the amount of 
squeezing and the quality of mode-matching to the atom are 
exactly the same for in-loop squeezing as for squeezing of a free 
field.  
Thus in-loop squeezing appears 
likely to be an important tool for future experimental investigation 
of the effect of low-noise light on atoms, as it is usually 
easier to generate than free squeezing.

The remainder of this letter is structured as follows.  First I  
review the theory of in-loop squeezing. Then I show how its effect on 
an in-loop atom can be described using quantum trajectories, including 
an arbitrary detector efficiency 
$\varepsilon$. In the limit of broad-band feedback
 I derive a master equation for the atom.  
 From this the spectrum of the fluorescence of the 
atom into the other (non-squeezed) radiation modes is easily 
calculated.  Then I compare these results with that obtained from 
broad-band free squeezing.  Finally, I discuss 
experimental implications of the theory.

{\em In-loop Squeezing.} 
Consider the experimental apparatus shown in Fig.~1, 
but for the moment without the fluorescent atom. 
The Mach-Zehnder interferometer on the left 
hand side has two functions. First it produces a weak beam 
($b_{\rm in}$ which is given by 
\beq
 b_{\rm in}(t) = \nu(t) -(i/2)(\beta e^{i\phi} - \beta e^{-i\phi})
\approx \nu(t) + \beta \phi.
\eeq
Here $\beta$, assumed real, is the coherent amplitude of the laser, 
and $\pm \phi$, assumed small, are the phase shifts imposed by the 
electro-optic modulators. The operator $\nu(t)$ 
represents vacuum fluctuations, for which all first and second order 
moments vanish except for \cite{Gar91}
\beq
\ip{\nu(t)\nu\dg(t')} = [\nu(t),\nu\dg(t')] = \delta(t-t').
\eeq
The second function of the interferometer is to produce a local 
oscillator beam with mean amplitude 
\beq
-(1/2)(\beta e^{i\phi} + \beta e^{-i\phi}) = -\beta[ 1 +O(\phi^{2})].
\eeq
With appropriate phase shifts assumed, this local oscillator is then 
used for making a homodyne measurement of the $X=b+b\dg$ quadrature of 
$b_{\rm out}$ (which, in the absence of the atom, is identical with 
$b_{\rm in}$). If the efficiency of the detectors is $\ve$ then the 
homodyne photocurrent is represented by the operator \cite{Gar91}
\beq \label{homI}
I_{\rm hom}(t) = \rt{\ve}X_{\rm out}(t)
+ \rt{1-\ve}\xi_{\ve}(t),
\eeq
where $\xi_{\ve}(t)$ is a unit-norm real white noise process.

The photocurrent \erf{homI} is normalized so that the spectrum of 
the stochastic process $I_{\rm hom}(t)$ (assumed to be stationary and of 
zero mean)
\beq
S^{X}_{\rm hom}(\omega) = \ip{\tilde{I}_{\rm hom}(\omega)I_{\rm hom}(0)}
\eeq
equals unity for $\omega \to \infty$. Here the tilde denotes the usual 
Fourier transform. The spectrum for $X_{\rm out}$ itself is defined 
analogously and is also given by 
\beq
S^{X}_{\rm out}(\omega) =
 \frac{1}{2\pi}\int d\omega' \ip{\tilde{X}_{\rm out}(\omega)\tilde{X}_{\rm out}(-\omega')}.
\eeq
At very high frequencies 
$\tilde{X}_{\rm out}(\omega)$ is dominated by the vacuum fluctuations 
$\xi_{\nu}(t) = \nu(t) + \nu\dg(t),$
which is a noise term like $\xi_{\ve}(t)$. In the Fourier domain these obey
\beq
\ip{\tilde{\xi}_{a}(\omega)\tilde{\xi}_{b}(-\omega')} = 2\pi\delta_{a,b}
\delta(\omega - \omega'),
\eeq
as required for the limit 
$S^{X}_{\rm out}(\infty)=S^{X}_{\rm hom}(\infty) = 1$. 

Although $S^{X}_{\rm hom}(\omega)$ and $S^{X}_{\rm out}(\omega)$ are 
shot-noise limited for high frequencies, they need not be for lower 
frequencies. In particular, the feedback loop shown can produce a 
spectrum below  the shot-noise as follows. The current $I_{\rm hom}(t)$ 
is amplified and used to control $\phi$. If we set 
\beq \label{setphi}
\phi(t) = \frac{g}{2\beta\sqrt{\ve}}\int_{0}^{\tau}h(s) I_{\rm 
hom}(t-s) ds,
\eeq
where $\tau > 0$ and $h(s)$ is normalized such that $\int_{0}^{\tau}h(s)ds=1$, 
then we have a feedback loop 
with a low-frequency round-loop gain of $g$ 
which is stable as long as $g{\rm Re}[\tilde{h}(\omega)]<1$ for all 
$\omega$. Solving in the Fourier domain,
\beq
\tilde{X}_{\rm out}(\omega) =
\left[{\tilde{\xi}_{\nu}(\omega) + 
g\tilde{h}(\omega)\rt{\frac{1-\ve}{\ve}}\tilde{\xi_{\ve}}(\omega)}\right]\frac{1}
{1-g\tilde{h}(\omega)}.
\eeq 
Thus $X_{\rm in}$ (which here equals $X_{\rm out}$) has a spectrum
\beq \label{spec}
S^{X}_{\rm in}(\omega) = 
[{1+g^{2}|\tilde{h}(\omega)|^{2}(\ve^{-1}-1)}]/{|1-g\tilde{h}(\omega)|^{2}}.
\eeq
At a frequency $\bar\omega$ much less than the feedback bandwidth $\sim 
\tau^{-1}$, $\tilde{h}(\bar\omega) = 1$ and the minimum noise is 
\beq \label{minin}
S^{X}_{\rm in}(\bar\omega)_{\rm min} = 1-\ve \;,\;\; {\rm for}\; 
g= - \ve/(1-\ve),
\eeq
which is clearly below the standard quantum limit.

Note that the condition to minimize $S^{X}_{\rm in}(\bar\omega)$ is not the 
same as that to minimize the  in-loop photocurrent noise:
\beq
S^{X}_{\rm hom}(\bar\omega)_{\rm min}\to 0 \;,\;\; {\rm as}\; g\to-\infty.
\eeq
It might be thought that this is the more relevant quantity, since 
$S^{X}_{\rm in}(\omega)$ is not actually measured in the experiment. 
However, a perfect QND \cite{WalMil94} 
device for measuring $X_{\rm in}$ would 
produce the spectrum (\ref{spec}) \cite{Sha87}. It is thus expected 
that for an in-loop atom, the 
relevant spectrum would again be $S^{X}_{\rm in}(\omega)$.

{\em In-loop Atom.}
Returning to Fig.~1, we now include the two-level atom, 
which is assumed to be resonant to the laser. It couples strongly 
only to modes of the radiation field having the appropriate dipole spatial 
distribution \cite{Gar91}. 
However, by focusing a beam as shown in Fig.~1, it is 
possible to mode-match a significant proportion, say $\eta$, of 
$b_{\rm in}$ into the atom's input.  In practice, a more efficient way 
to increase the effective $\eta$ would be to couple the light into a
 microcavity, as in Ref.~\cite{Ver98}. The Hamiltonian of the atom in 
the interaction picture at time $t$ is then
\beq  \label{Hfun}
H(t) = -i[\rt\eta b_{\rm in}(t) + 
\rt{1-\eta}\mu(t)]\sigma\dg(t) + {\rm H.c.}
\eeq
Here $\sigma = \ket{g}\bra{e}$ is the atomic lowering operator and
I have set the longitudinal atomic decay rate to unity. The 
operator $\mu(t)$ represents an independent vacuum input. Under this 
coupling, the output field is given by \cite{Gar91}
\beq \label{ior}
b_{\rm out}(t) = b_{\rm in}(t) + \rt\eta \sigma(t).
\eeq

Although it would be possible to give a description of the entire 
feedback loop in terms of atomic and radiation field operators, it is 
simpler to use the quantum trajectory theory of homodyne measurement 
\cite{Car93b,WisMil93a}. In this theory, only the atom is treated as 
a quantum mechanical system with state matrix $\rho(t)$; 
the rest of the apparatus is considered 
as a complicated measurement and feedback device for the atom. The 
photocurrent $I_{\rm hom}(t)$ is therefore a classical quantity. It 
is given by  
\beq \label{ce1}
I_{\rm hom}(t) = \bar{I}_{\rm hom}(t) + \xi_{\rm hom}(t).
\eeq
where $\xi_{\rm hom}(t)$ is local-oscillator shot noise, which in this 
theory is the only source of noise in the whole system. 
From Eqs.(\ref{homI}) and (\ref{ior}), the expected value 
$\bar{I}_{\rm hom}(t)$ is 
\beq \label{ce2}
\bar{I}_{\rm hom}(t) = \rt{\eta\ve} {\rm Tr}[\rho(t)\sigma_{x}] + 
\rt{\ve}2\beta\phi(t).
\eeq
Here $\phi(t)$ is not set to its average value of zero because 
it is, in principle, known at any time as it is 
determined by the prior classical photocurrent via \erf{setphi}. 

In the quantum trajectory theory, the photocurrent  noise 
directly affects the atom, via a nonlinear 
stochastic term in the atom's master equation 
\cite{WisMil93a}
\beq
d\rho = dt{\cal D}[\sigma]\rho
 + \rt{\eta\ve}dW_{\rm hom}(t){\cal H}[\sigma]\rho
-idt[H_{\rm fb},\rho].
\eeq
The first term represents the usual damping, with
\beq
{\cal D}[A]B \equiv ABA\dg - \half A\dg A B - \half B A\dg A.
\eeq
The second term represents the conditioning by the measurement, with
$dW_{\rm hom}(t)=\xi_{\rm hom}(t)dt$ and
\beq
{\cal H}[A]B \equiv AB + BA\dg - {\rm Tr}[AB+BA\dg]B.
\eeq
The final Hamiltonian is due to the feedback. It is 
identical to the term due to feedback in the fundamental atomic 
Hamiltonian \erf{Hfun}, namely
\beq \label{sub1}
H_{\rm fb}(t) =  \rt{\eta} \beta\phi(t) \sigma_{y}.
\eeq

Now consider the limit of instantaneous feedback on the atomic 
time-scale, $\tau \ll 1$. In this limit 
\beq \label{sub2}
2\beta\phi(t) = gI(t)/\sqrt{\ve}
\eeq
and thus we can derive 
\beq
I_{\rm hom}(t) = (1-g)^{-1}\{\xi(t) + \rt{\eta\ve} 
{\rm Tr}[\rho(t)\sigma_{x}]\}.
\eeq
Hence from Eqs.~(\ref{sub1}) and (\ref{sub2}),
\beq
H_{\rm fb}(t) = \lambda \half \sigma_y \{
 {\rm Tr}[\rho(t)\sigma_{x}]+\xi(t)/\rt{\eta\ve}\},
\eeq 
where it is to be understood that $t$ on the right-hand side of 
this equation actually stands for $t-0^+$. The feedback parameter 
$\lambda$ is given by  
\beq
\lambda = {g\eta}/({1-g}) \; \in (-\eta,\infty).
\eeq

Thus far we still have a nonlinear stochastic equation 
for the conditioned atomic state matrix $\rho$, 
which is not easy to work with. 
However, in the Markovian limit it is possible 
to average over the stochasticity both in the measurement 
and feedback terms. 
As explained in Refs.~\cite{WisMil93b,WisMil94a}, the result 
is the master equation
\beq \label{cenme}
\dot\rho = {\cal D}[\sigma]\rho 
-i\lambda[\half\sigma_y,\sigma \rho + \rho\sigma\dg]
+ \frac{\lambda^2}{\eta\ve}{\cal D}[\half\sigma_y]\rho.
\eeq
This equation, and 
the following relation between $\lambda$ and the in-loop squeezing 
\beq \label{sbw}
S^{X}_{\rm in}(\bar\omega) = 
\frac{1+g^{2}(\ve^{-1}-1)}{(1-g)^{2}} 
= 1+\frac{2\lambda}{\eta}+\frac{\lambda^{2}}{\eta^{2}\ve}.
\eeq
are the central results of this work. In \erf{sbw}, we still have 
$\bar\omega \ll \tau^{-1}$, but now also 
 $\bar\omega\gg 1$. This ensures that the atomic variables 
(with characteristic time scale of unity) do not contribute 
significantly to the 
spectrum at $\bar\omega$, so that \erf{spec} is still valid. 

From the master equation \erf{cenme} it is easy to derive the 
following dynamical equations:
\bqa
{\rm Tr}[\dot\rho \sigma_{x}] &=& -\gamma_{x}{\rm Tr}[\rho 
\sigma_{x}] \label{onex}\\
{\rm Tr}[\dot\rho \sigma_{y}] &=& -\gamma_{y}{\rm Tr}[\rho \sigma_{y}]\\
{\rm Tr}[\dot\rho \sigma_{z}] &=& -\gamma_{z}{\rm Tr}[\rho 
\sigma_{z}]-C \label{threez}
\eqa
Only the equation for $\sigma_{y}$ is unaffected by the feedback, 
with $\gamma_{y} = 1/2.$  The new decay rate for $\sigma_{x}$ is
\beq
\gamma_{x} = \frac12\left[ 1+2\lambda + 
\frac{\lambda^{2}}{\eta\ve}\right],
\eeq
and the modified parameters for $\sigma_{z}$ are
\beq
\gamma_{z} = \gamma_{y} + \gamma_{x}\;,\;\;C = 1+\lambda,
\eeq
In steady state ${\rm Tr}[\rho_{\rm ss}\sigma_{x}]=
{\rm Tr}[\rho_{\rm ss}\sigma_{y}]=0$ and
${\rm Tr}[\rho_{\rm ss}\sigma_{z}] = -1 +
{\lambda^{2}}/[{2\eta\ve(1+\lambda)+\lambda^{2}}]$.

The most interesting of these results is that 
negative feedback can reduce 
the decay rate of the $x$ component of the atomic dipole 
below its natural value of $1/2$.
From \erf{sbw} it can be re-expressed as
\beq \label{gx1}
\gamma_{x} = \half \left[ (1 - \eta) + \eta S^{X}_{\rm 
in}(\bar\omega)\right] .
\eeq
This clearly shows that $\gamma_{x}$ has two contributions: 
$\half(1-\eta)$ from the vacuum input and $\half \eta S^{X}_{\rm 
in}(\bar{\omega})$ from the in-loop squeezed light.
The greatest 
reduction occurs for minimum in-loop fluctuations 
as in \erf{minin}, for which
\beq
(\gamma_{x})_{\rm min} = \half\left( 1- \eta\ve \right) \;,\;\;
{\rm for}\;\lambda = -\eta\ve.
\eeq

The slower decay of $\sigma_{x}$ can be directly observed in the 
power spectrum of the fluorescence of the atom into the vacuum modes. 
This measures the photon flux per unit frequency into these modes
and is defined by
\beq
P(\omega) = \frac{1-\eta}{2\pi}\ip{\tilde{\sigma}\dg(-\omega)
\sigma(0)}_{\rm ss}.
\eeq
From \erf{cenme} this is easily evaluated to be
\beq \label{genps}
P(\omega) = \frac{(1-\eta)(\gamma_{z}-C)}{8\pi\gamma_{z}}\left[ 
\frac{\gamma_{x}}{\gamma_{x}^{2}+\omega^{2}} + 
\frac{\gamma_{y}}{\gamma_{y}^{2}+\omega^{2}}\right].
\eeq
For the optimal squeezing ($\lambda =-\eta\ve$) we have
\beq
P(\omega) = \frac{(1-\eta)\eta\ve}{4\pi(2 - \eta\ve)} \left[ 
\frac{1-\eta\ve}{(1-\eta\ve)^{2}+4\omega^{2}} + 
\frac{1}{1+4\omega^{2}}\right].
\eeq
This is plotted in Fig.~2 for $\eta=0.8$ and $\ve=0.95$.

{\em Comparison with Free Squeezing.} 
To compare the above results with those produced by free squeezing we 
again assume that the mode-matching of the squeezed modes into the atom is 
$\eta$, and that the squeezing is broad-band compared to the atom. 
Assuming also that the input light is in a 
minimum-uncertainty state for the $X$ and $Y$ quadratures 
\cite{WalMil94}, 
it can be characterized by a single real number $L$, with
\beq\label{L}
S_{\rm in}^{X}(\omega) = L = 1/S_{\rm in}^{Y}(\omega) .
\eeq
In conventional notation \cite{Gar91}, $L=2N+2M+1$,
where $M^{2}=N(N+1)$. This yields the master equation
\beq
\dot\rho = (1-\eta){\cal D}[\sigma]\rho + \frac{\eta}{4L}
{\cal D}[(L+1)\sigma-(L-1)\sigma\dg]\rho,
\eeq
which leads again to Eqs.(\ref{onex})--(\ref{threez}), but with
\bqa
\gamma_{x} &=& \half\left[ (1-\eta)+\eta L\right] \label{gx2},\\
\gamma_{y} &=& \half\left[ (1-\eta)+\eta L^{-1}\right], \\
\gamma_{z} &=& \gamma_{x}+\gamma_{y} \;,\;\;C=1.
\eqa

For $L<1$ the decay of  $\sigma_{x}$ is again inhibited.  
The crucial observation to be made is that the dependence of 
$\gamma_{x}$ on the degree of $X$ quadrature squeezing of the input 
light is exactly the same as for in-loop squeezing, as is seen by 
comparing Eqs.~(\ref{L}) and (\ref{gx2}) with \erf{gx1}. The only  
difference between the two cases is that $C$ is unaffected by the 
free squeezing and that $\gamma_{y}$ is not increased by the in-loop 
squeezing. The latter is a direct consequence of the fact that an 
in-loop field is not bound by the usual two-time uncertainty relations. The 
free squeezing fluorescence spectrum is again given by \erf{genps}.
This is also plotted in Fig.~2 for $\eta=0.8$ and $L=0.05$. 
As this figure shows, the spectra are certainly not identical, 
but the sub-natural linewidth is much the same in both.

To conclude, line-narrowing of an 
atom is not a diagnostic of free squeezing.  Rather, it requires 
only temporal anticorrelations of one 
quadrature of the input field (for times much shorter than the 
atomic lifetime) such as can be produced by a negative electro-optic 
feedback loop.  The dependence of the line-narrowing on the input 
squeezing and the degree of mode-matching is the same for in-loop 
squeezing as for free squeezing.  Because the quadrature operators of 
an in-loop field do not obey the usual two-time commutation relations, 
the reduction in noise in one quadrature does not imply an increase in 
noise in the other.  Hence the line-narrowing of one quadrature of the 
atomic dipole by in-loop squeezing does not entail the line-broadening 
of the other quadrature. What significance this difference has in 
the physics of more complex atomic interactions with squeezed light 
\cite{FicDru97} is a question requiring much investigation.

In-loop squeezing is generally easier to produce than free squeezing 
for a number of reasons.  First, in-loop squeezing does not require 
expensive and delicate sources such as nonlinear crystals, but rather 
off-the-shelf electronic and electro-optical equipment.  Second, the 
amount of squeezing is limited only by the efficiency of the 
photodetection.  For homodyne detection, as required here, an 
efficiency of 95\% is readily obtainable \cite{Sch96} and would enable in-loop 
squeezing of 95\%.  Third, in-loop squeezing can be produced 
at any frequency for which a 
coherent source is available, so experiments could be conducted on any 
atomic transition.  The one difficulty with in-loop squeezing is that 
it requires a feedback loop response time much shorter than an atomic 
lifetime, but this would not be a problem for metastable transitions.  
Thus as well as  giving us a better theoretical understanding 
of the effects on matter of light with 
fluctuations below the standard quantum limit, in-loop squeezing  
should be a practical alternative to free 
squeezing in the experimental investigation of these effects.

{\em Acknowledgments.} 
I would like to thank G. Toombes for constructive criticisms. 
This work was supported by the Australian Research Council.

\begin{figure}
	\vspace{4.5cm}
	\begin{center}
		\special{illustration 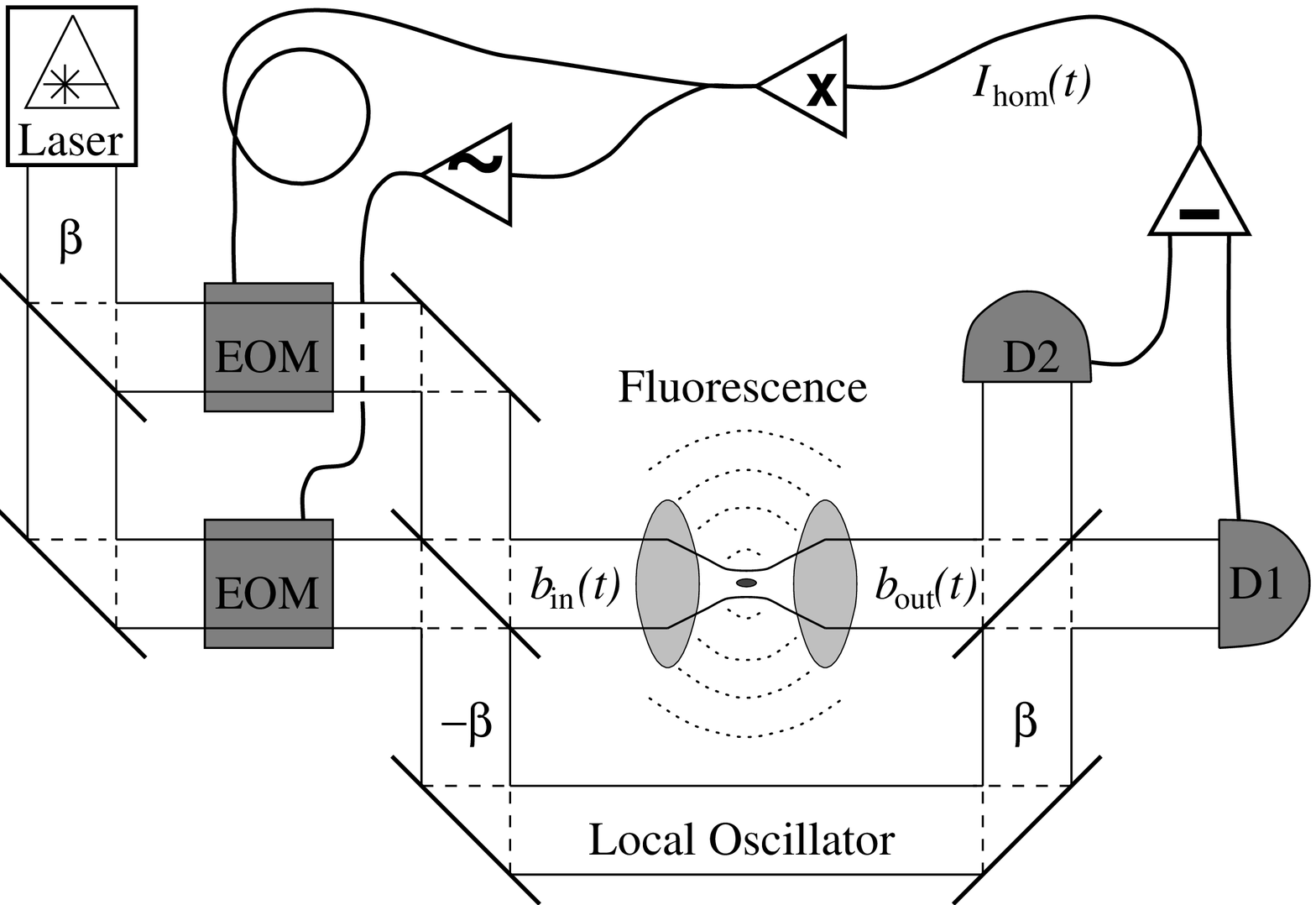 scaled 0.35}
	\end{center}
	\vspace{-0.5cm}
	\caption{\narrowtext Diagram of the experimental configuration 
	discussed. All beam splitters are 50:50. 
	The atom is represented by the small ellipse at 
	the focus of $b_{\rm in}(t)$. The difference $I_{\rm hom}(t)$ 
	between the photocurrents at detectors D1 and D2 is  amplified and 
	split. The two signals (with opposite sign) are 
	fed back to the two electro-optic  	modulators (EOM). 
	}
	\protect\label{fig1}
\end{figure}

\begin{figure}[h!]
	\vspace{5.5cm}
	\begin{center}
		\special{illustration 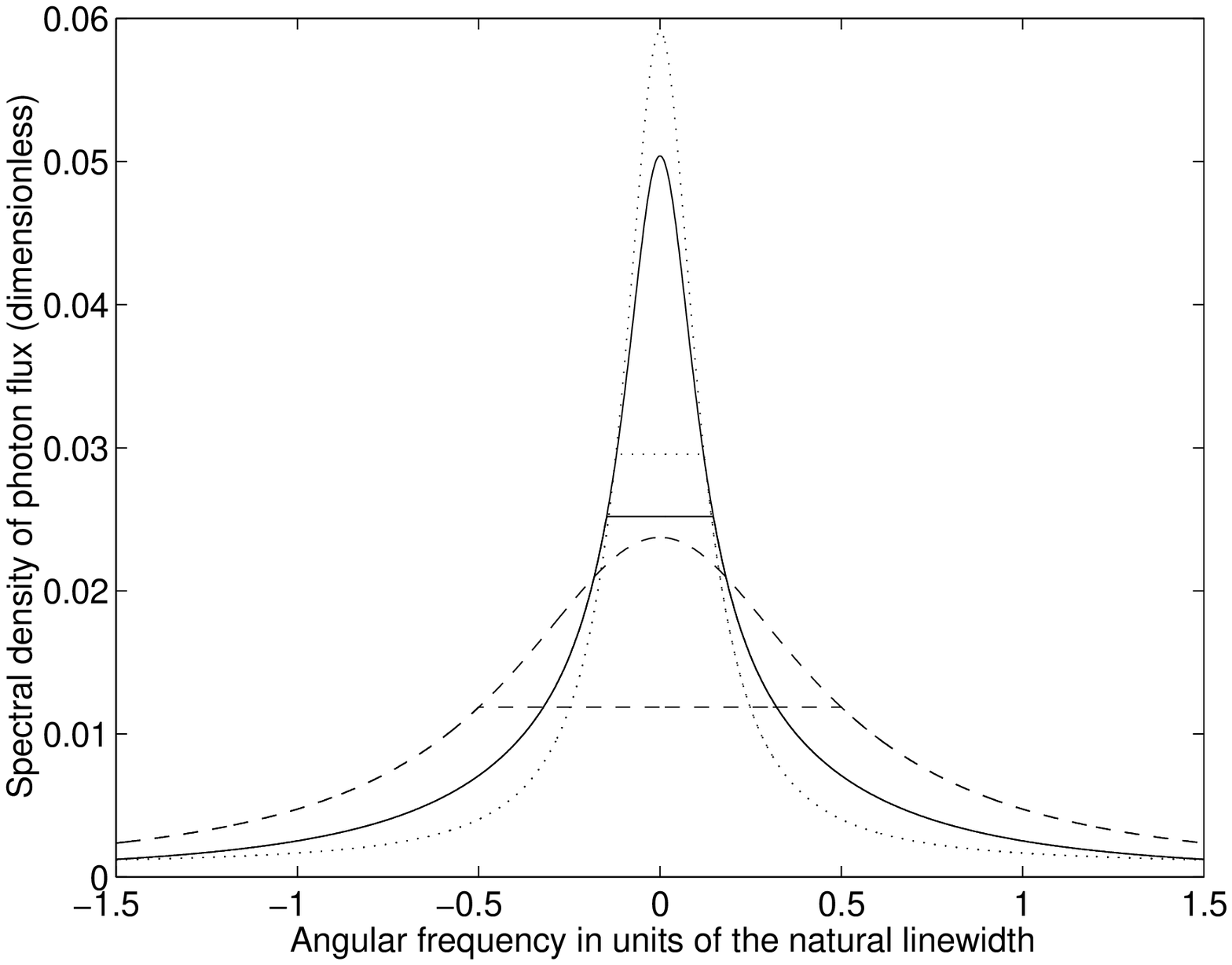 scaled 0.45} 
	\end{center}
	\vspace{-0.5cm}
	\caption{\narrowtext Plot of the Power Spectrum $P(\omega)$ of the 
	fluorescence into the vacuum modes, for  in-loop squeezing 
	(solid) and free squeezing (dotted), with 
	mode-matching $\eta=0.8$ and squeezing 
	$S^{X}_{\rm in}(\bar\omega)=0.05$. The linewidth for  
	in-loop squeezing is slightly broader  
	because the contribution from $\sigma_{y}$ 
	is not broadened in this case. The  
	natural-width spectrum of a very weakly driven atom (dashed) 
	is scaled up for comparison. }
	\protect\label{fig2}
\end{figure}

\end{multicols}

\end{document}